\documentclass[a4paper,10pt,twoside]{cpc-hepnp}

\usepackage{multicol}
\usepackage{graphicx}
\usepackage{booktabs}
\usepackage{amssymb,bm,mathrsfs,bbm,amscd}
\usepackage[tbtags]{amsmath}
\usepackage{lastpage}

\begin{document}

\fancyhead[c]{\small Submitted to Chinese Physics
C¡±} \fancyfoot[C]{\small 010201-\thepage}

\footnotetext[0]{Received 14 March 2009}

\title{Prototype of Readout Electronics for the ED in LHAASO KM2A\thanks{Supported by National Natural Science
Foundation of China (11375210) and the Knowledge Innovation Fund of IHEP, Beijing}}

\author{%
      LIU Xiang(ÁõÏæ)$^{1,2,3;1)}$\email{liuxiang@ihep.ac.cn}%
\quad CHANG Jing-Fan(³£¾¢·«)$^{1,2,3;2)}$\email{changjf@ihep.ac.cn}%
\quad WANG Zheng(Íõï£)$^{1,2,3}$
\quad FAN Lei(·®ÀÚ)$^{1,2,3}$
}
\maketitle

\address{%
$^1$ State Key Laboratory of Particle Detection and Electronics,Chinese Academy of Sciences,Beijing 100049,China\\
$^2$ Institute of High Energy Physics, Chinese Academy of Sciences, Beijing 100049, China\\
$^3$ University of Chinese Academy of Sciences, Beijing 100049, China\\
}

\begin{abstract}
The KM2A(one kilometer square extensive air shower array) is the largest detector array in the LHAASO(Large High Altitude Air Shower Observatory) project. The KM2A consists of 5635 EDs(Electromagnetic particle Detectors) and 1221 MDs(Muon Detectors). The EDs are distributed and exposed in the wild. Two channels, Anode and Dynode, are employed for the PMT(photomultiplier tube) signal readout. The readout electronics proposed in this paper aims at the accurate charge and arrival time measurement of the PMT signals, which cover a large amplitude range from 20P.E(photoelectrons) to 2$\times 10^5$P.E. By using the ¡°Trigger-less¡± architecture, we digitize signals close to the PMTs. All digitized data is transmitted to DAQ(Data Acquisition) via the simplified WR(White Rabbit) protocol. Compared with traditional high energy experiments, high-precision of time measurement in such a large area and suppression of temperature effects in the wild become the key techniques. Experiments show that the design has fulfilled the requirements in this project.
\end{abstract}



\footnotetext[0]{\hspace*{-3mm}\raisebox{0.3ex}{$\scriptstyle\copyright$}2013
Chinese Physical Society and the Institute of High Energy Physics
of the Chinese Academy of Sciences and the Institute
of Modern Physics of the Chinese Academy of Sciences and IOP Publishing Ltd}%

\begin{multicols}{2}

\section{Introduction}
The LHAASO (Large High Altitude Air Shower Observatory)\cite{lab1} is a multi-objective project aiming at searching the origin of high energy galactic cosmic rays with a $1.2km^2$ complicated ground detector array. The KM2A (one kilometer square extensive air shower array), designed for the measurement of the number density and arrival time of shower particles, contains 5635 EDs (Electromagnetic particle Detectors) and 1221 MDs (Muon Detectors)\cite{lab2}. All EDs are evenly distributed in the one square kilometer wild area.

Signals from PMTs vary in a large dynamic range from 20P.E(photoelectrons)(single particle) to 2$\times 10^5$P.E($10^4$ particles)\cite{lab3}, because of the large primary energy of the cosmic rays(10 TeV-100 TeV) and the different distances between detectors and shower cores. To cover the whole dynamic energy range and guarantee the accurate case reconstruction, precise charge and time measurements should be achieved in readout electronics. In this broad area, we need to read out 5635 PMTs in total. As for charge measurement, a resolution of 25$\%$ at single particle is required for each channel. And the RMS(root mean square) of time measurement is demanded to be better than 1ns(nanosecond) while the bin size is better than 2ns, which means that the high quality of clock synchronization and distribution in a large scale is required. Furthermore, the performance stability become a great challenge because the electronics will be running in the wild and the temperature can vary from $-$5$^\circ$C to $+$50$^\circ$C in the high altitude observatory. To sum up, the requirement for electronics is list in Table~\ref{tab1}.
\end{multicols}

\begin{center}
\tabcaption{ \label{tab1}  Requirement of the ED readout electronics}
\footnotesize
\begin{tabular*}{170mm}{@{\extracolsep{\fill}}ccccccc}
\toprule $Item$ & $Requirement$\\
\hline
$Dynamic\hphantom{0}range\hphantom{0}of\hphantom{0}charge\hphantom{0}measurement$\hphantom{00} & \hphantom{0}1-10000particles\\
$Resolution\hphantom{0}of\hphantom{0}charge\hphantom{0}measurement$\hphantom{00} & \hphantom{0}25$\%$@1particle;\hphantom{0}5$\%$@10000particle\\
$Bin\hphantom{0}size\hphantom{0}of\hphantom{0}time measurement$\hphantom{00} & \hphantom{0} \textless 2ns\\
$Resolution\hphantom{0}of\hphantom{0}time\hphantom{0}measurement$\hphantom{00} & \hphantom{0} \textless 1ns\\
$Case\hphantom{0}rate\hphantom{0}when\hphantom{0}threshold\hphantom{0}is\hphantom{0}set\hphantom{0}0.125 particle$\hphantom{00} & \hphantom{0}2KHz\\
$Temperature\hphantom{0}range$\hphantom{00} & \hphantom{0} $-$5$^\circ$C -- $+$50$^\circ$C\\
\bottomrule
\end{tabular*}%
\end{center}

\begin{multicols}{2}
In the view of the distributed architecture of detectors, the traditional electronics system based on VME case is significantly unpractical. Instead, we adopt a trigger-less, separate, front-end architecture, in which we place electronics nearby the PMT to achieve the purpose of maintaining signal characteristics and reducing the number of signal cables. Digitized data, commands and clocks are transferred via a single optical fiber.

\begin{center}
\includegraphics[width=6cm]{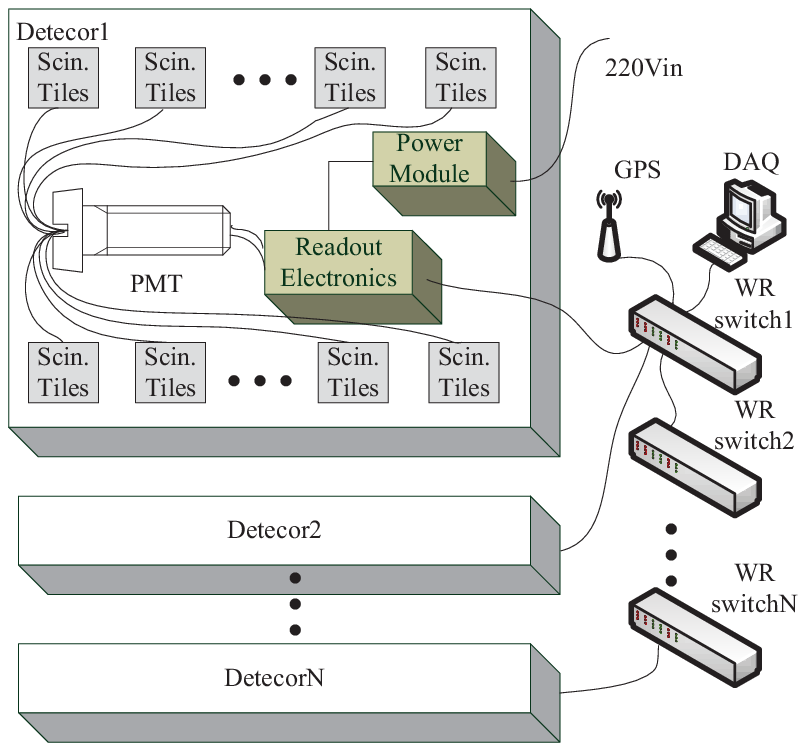}
\figcaption{\label{fig1}   Architecture of EDs}
\end{center}

Fig.~\ref{fig1} shows the structure of EDs. The readout electronics, power module are placed in the same shell with tiles and PMT. The 220V AC (alternating current) power is modulated to ¡À6V DC(direct current) supply for the electronics and settable high-voltage supply for PMT. Each ED consists of 16 scintillator tiles which are readout by wavelength-shifting fibers bundled and fixed by a 25mm PMT\cite{lab4}. Signals from PMT are transmitted to the electronics via two independent 50cm cables.

The local clocks of ED readout electronics are synchronized by a global high precision 125MHz clock, which is created and distributed based on the simplified WR(White Rabbit) protocol\cite{lab5}. This protocol is proposed by CERN and simplified by Tsinghua University, and developed to provide a sub-nanosecond accuracy and picosecond precision of synchronization. The synchronized clock can be distributed to thousands of WR nodes within 10km, while the timing and data link are combined over the same physical media. Experiments show that clocks of two WR nodes can be synchronized with 100ps(picoseconds) accuracy and 20ps precision\cite{lab6}\cite{lab7}. According to WR protocol, a CUTE-WR(compact universal timing endpoint based on the WR) is required in the timing synchronization. In this system, the CUTE-WR hardware is fused into the readout electronics. Function of CUTE-WR is recovering the TAI
(international atomic time), PPS(pulse per second) and synchronized 125MHz clock. Besides, a standard network interface is supplied for data transmission.

Compared with the architecture of global trigger in traditional experiments such as BESIII\cite{lab8} and LAWCA\cite{lab9}, ED readout electronics needs no trigger electronics. Data discriminated by the threshold is transmitted to DAQ(Data Acquisition) and the kernel trigger processing is also completed in DAQ. The TUTE-WR contains a programmable core named WRPC(WR PTP core)\cite{lab10}, which can serve as an Ethernet MAC in FPGA. In order to improve the reliability of data transmission, an implementation of TCP/IP protocol based on FPGA, cooperating with WRPC has been developed to forward and receive data packages between readout electronics and DAQ. The results tell that the maximum throughput of 320Mbits per second can be achieved in this realization\cite{lab11}, that is sufficient for the ED readout system due to the highest case rate of PMT is 2KHz when the threshold is set to 1/8 particle.
\section{The ED Readout Electronics}
The architecture of ED readout electronics is designed and measured as shown in Fig.~\ref{fig2}. Signals from Anode and Dynode are digitized by a two-channels, 65MHz ADC(Analog to Digital Converter) after handled by the amplification and shaping circuit. Then, the digitized signals is sent to FPGA for digital peaking. Signals from Anode are also amplified and compared with the threshold to trigger time measurement. This trigger is delivered to FPGA and then digitized by a TDC (time to digital converter) integrated in the FPGA.

\begin{center}
\includegraphics[width=8cm]{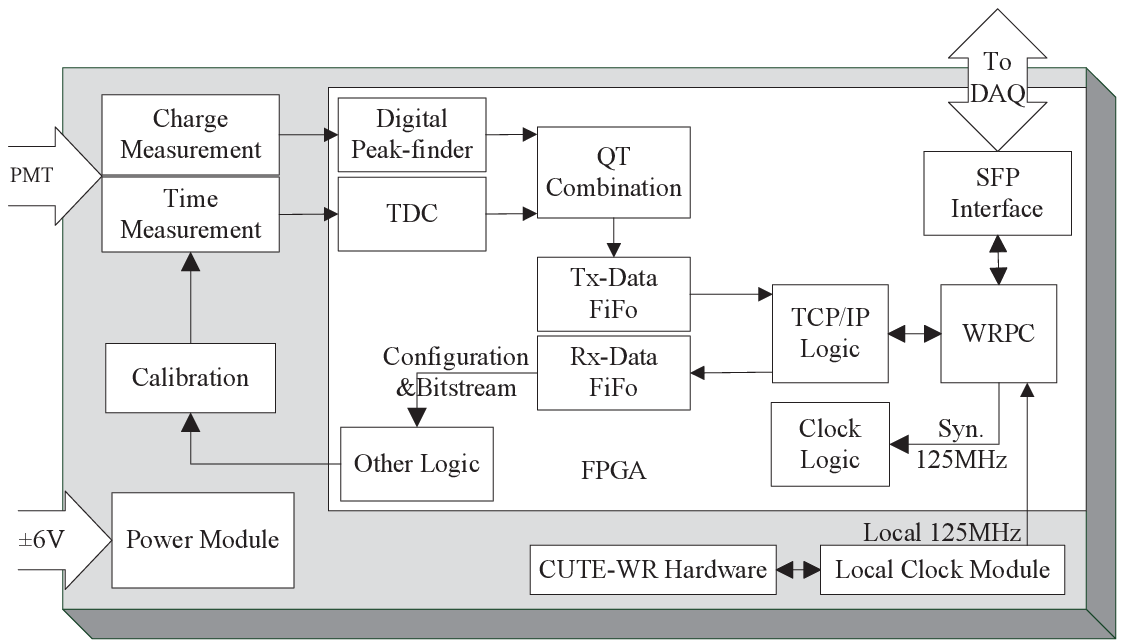}
\figcaption{\label{fig2}   Block diagram of ED readout electronics}
\end{center}

Digitized charge and time information is combined and buffered in Tx-FIFO. The TCP/IP module read and send the data in Tx-FIFO in each millisecond. Configuration data, control commands and the bitstream files is received and buffered in the Rx-FIFO. The WRPC adjust local 125 MHz clock provided by onboard crystal oscillator to a synchronized referent clock, which can be regulated to 62.5MHz clock for system process and 250MHz clock for TDC module.

\subsection{Charge Digitizing Module Design}
To achieve the large scope as mentioned and get a enough overlap, two electronics channels and two PMT readout channels are employed. The Anode electronics channel covers a range of 1$\sim$200 particles while the Dynode channel covers a range of 100$\sim 10^4$ particles. And the ratio of gain between two channel is designed 2 times.

The charge digitizing circuit is shown in Fig.~\ref{fig3}. To achieve the high precision of charge measurement and better temperature performance, this circuit abandon the input buffer compared with the readout electronics for WCDA(water Cherenkov detector array)\cite{lab12} in LHAASO. Simulation and experiments shows that the offset voltage of the buffer amplifier is obviously amplified by integral circuit.So, a little swing of this offset voltage can literally deteriorate the SNR(signal-noise ratio) of the charge measurement. What¡¯s more, the amplified offset voltage is tested to vary evidently with the temperature shift.

\begin{center}
\includegraphics[width=8cm]{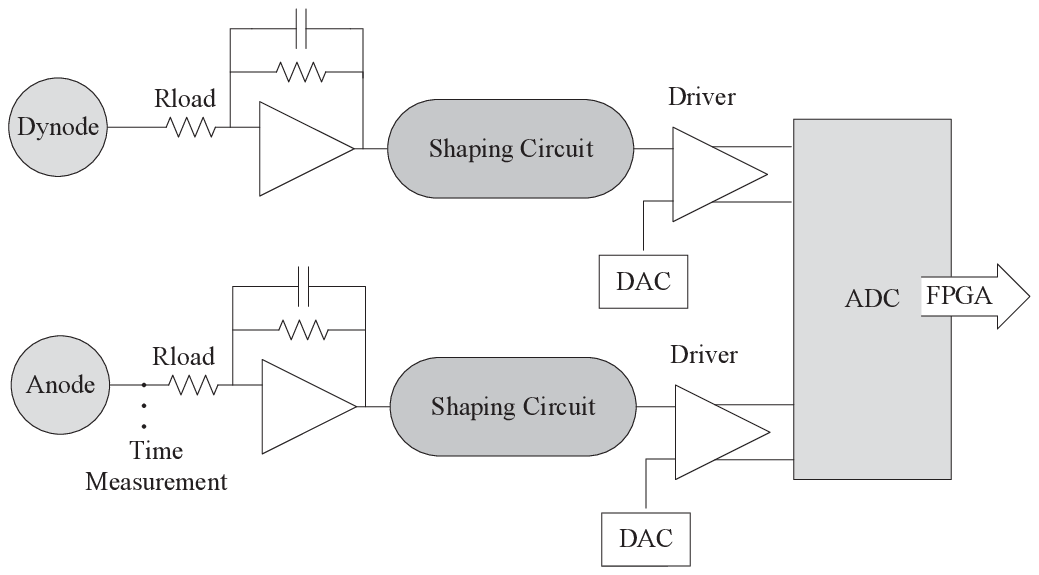}
\figcaption{\label{fig3}   Design of charge digitizing module}
\end{center}

In this design, current signals from PMTs are directly sent to integral and shaping circuit. The peak value digitized by ADC and sought in FPGA can quantitatively represent the charge of input signal. The gain resistor in the integral circuit is devised to be 50Ohm to serve as an input load resistor. Voltage signals for time measurement is also sampled at the gain resistor of anode channel.

To reduce the effect of temperature, the DC performance of the amplifiers becomes significant. Amplifiers with lower input offset voltage drift are selected in this design. By the way, the inverting input of the ADC differential driver is supposed to source from low drift referent DAC(digital to analog converter) instead of mechanical potentiometer.

\subsection{Time Digitizing Module design}
The technology of Multi-phase TDC\cite{lab13} based on FPGA is applied to measure arrival time of Anode signal, as illustrated in Fig.~\ref{fig4}. Voltage sampled from Anode is magnified 20 times by two steps. We assign the gain of the first step to be +2 for the input high-impedance state, therefore, the input impedance of Anode keeps 50Ohm.

\begin{center}
\includegraphics[width=8cm]{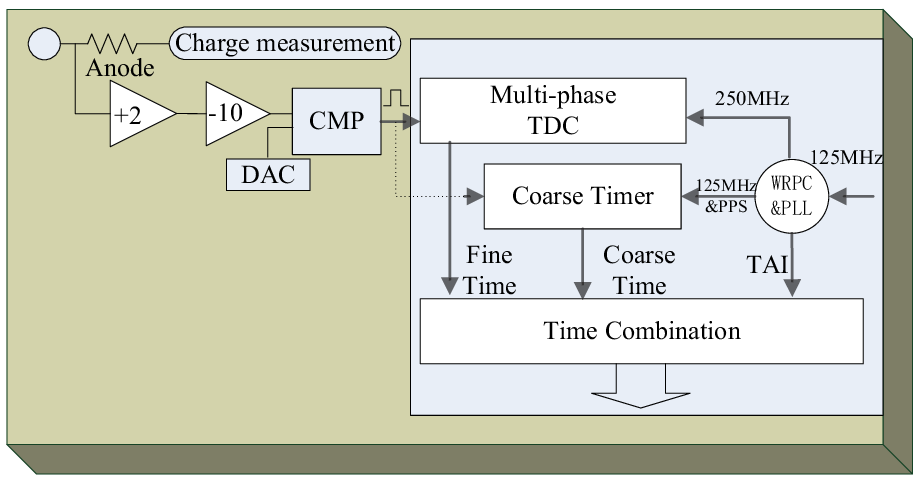}
\figcaption{\label{fig4}   Design of time digitizing module}
\end{center}

In consideration of the waveform of the PMT signal, the voltage sampled from the 50Ohm resistor is about 8mV when the Anode input signal equal single particle. A 2.5V full-scale, 12bits DAC is employed as a threshold generator. Since the step of DAC is 0.61 mV, threshold of time measurement can accurately be set to 1/8 particle.

The time digitizing module integrated in the FPGA is triggered by the rising edge from comparator. All clocks related in this section originate from the synchronized 125MHz. The coarse time counter is driven by a 125 MHz clock and reset by the PPS signal, so the dynamic range of time measurement is 1s. Four 250MHz phase-shifted clocks drive the fine time measurement, the bin size of the TDC is 1ns in this way. Finally, the TAI time from WRPC, coarse time from coarse counter and fine time from multi-phase circuits constitute the arrival time of signals.

The lower input offset voltage drifts of amplifiers and DAC contribute to the better temperature performance. On the other hand, the routing length of trigger signals from comparator to the first flip-flop in the FPGA should be controlled because the routing delay is also proved to drift with temperature changing.

\section{Performances of the ED Readout Electronics}
\subsection{Charge Digitizing Module evaluation}
To evaluate the charge measurement in laboratory, we set the frequency of generated signal to the maximum event-ratio 2kHz. And the peak value of anode signals is adjusted from 8mV to 1600mV to cover the full range. Correspondingly, the dynode signals is adjusted from 4mV to 700mV.

Resolution of the minimum charge measurement is shown in Fig.~\ref{fig5}, where Mean and RMS are
the mean value and standard deviation of signals. According to the left figure, the resolution of Dynode charge measurement is calculated to 2.19\%, and a 4.88\% precision of anode can be deducted from right figure.

\end{multicols}
\begin{center}
\includegraphics[width=12cm]{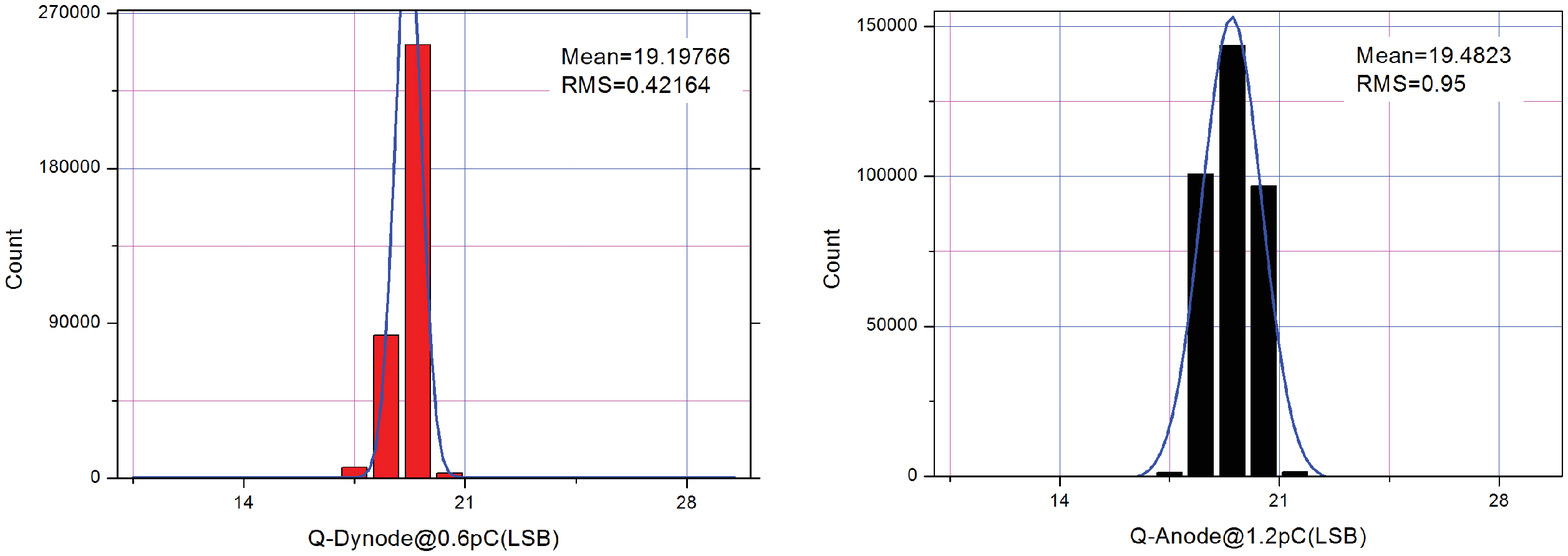}
\figcaption{\label{fig5}   Precision of the minimum signals }
\end{center}
\begin{multicols}{2}

\end{multicols}

\begin{center}
\tabcaption{ \label{tab2}  Contrast result of precision}
\footnotesize
\begin{tabular*}{170mm}{@{\extracolsep{\fill}}ccccccc}
\toprule $$ & $Anode$  & $Anode(experimental)$ & $Dynode$ & $Dynode(experimental)$ \\
\hline
$Mean(LSB)$\hphantom{00} & \hphantom{0}19.48 & \hphantom{0}22.57 & 19.20 & 24.07\\
$RMS(LSB)$\hphantom{00} & \hphantom{0}0.95 & \hphantom{0}3.01 & 0.42 & 4.17\\
$Precision$\hphantom{00} & \hphantom{0}4.88\% & \hphantom{00}13.33\%\hphantom{00} & 2.19\%\hphantom{0} & 17.32\%\\
\bottomrule
\end{tabular*}%
\end{center}

\begin{multicols}{2}

To verify the influences of input buffer on the precision as described, we similarly test the performance of experimental board which is equipped with input buffers both in dynode and anode channel. Table~\ref{tab2} tells the contrast result. Although the precision can also fulfill the requirement of 25\%, the input buffers clearly worsen the performances of charge measurement.

\end{multicols}
\begin{center}
\includegraphics[width=12cm]{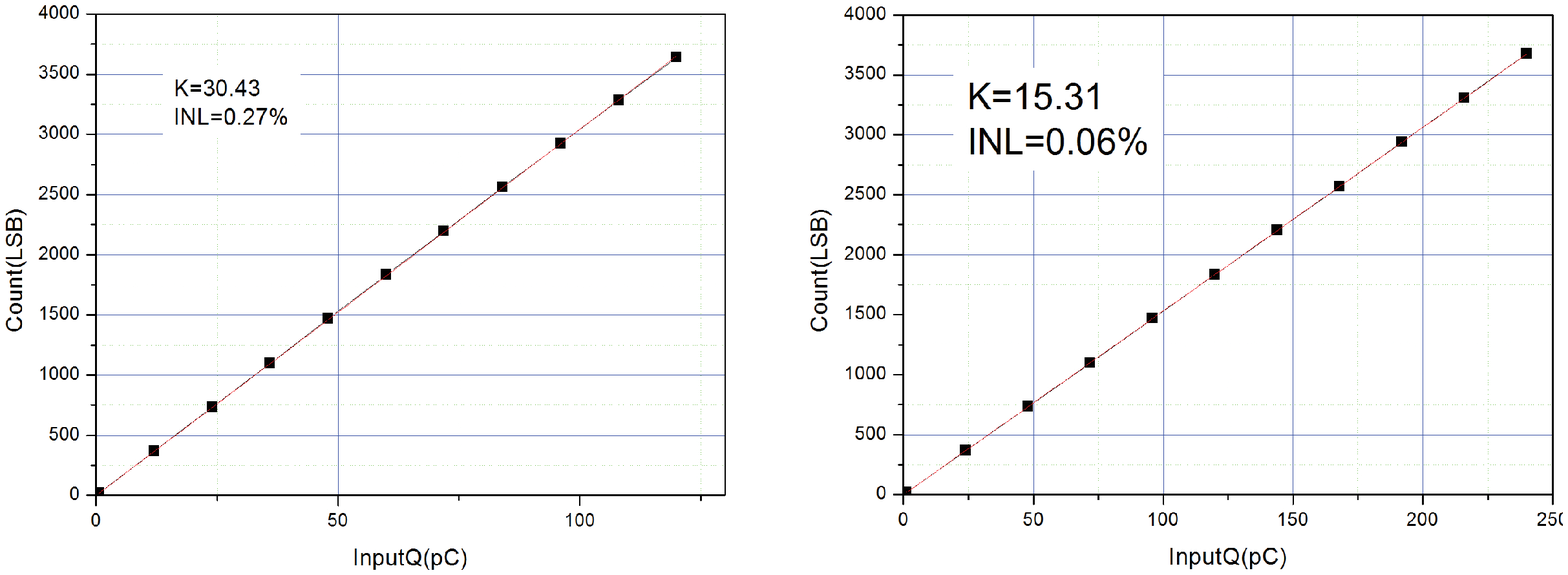}
\figcaption{\label{fig6}   INL of charge digitizing module }
\end{center}
\begin{multicols}{2}

The INL(integral non-linearity) of charge digitizing module is tested as shown in Fig.~\ref{fig6}. The K means the gain of ADC readout to input charge. The proportion between dynode and anode is proved to be 1.98.

\subsection{Time Digitizing Module evaluation}
A statistical code density test based on a source of random hits\cite{lab13} is used to evaluate the INL of time digitizing module. Since the coarse counter is driven by 125MHz clock, the fine time is divided to 8 bins. The maximum INL achieved 0.03LSB as shown in Fig.~\ref{fig7}.

\begin{center}
\includegraphics[width=6cm]{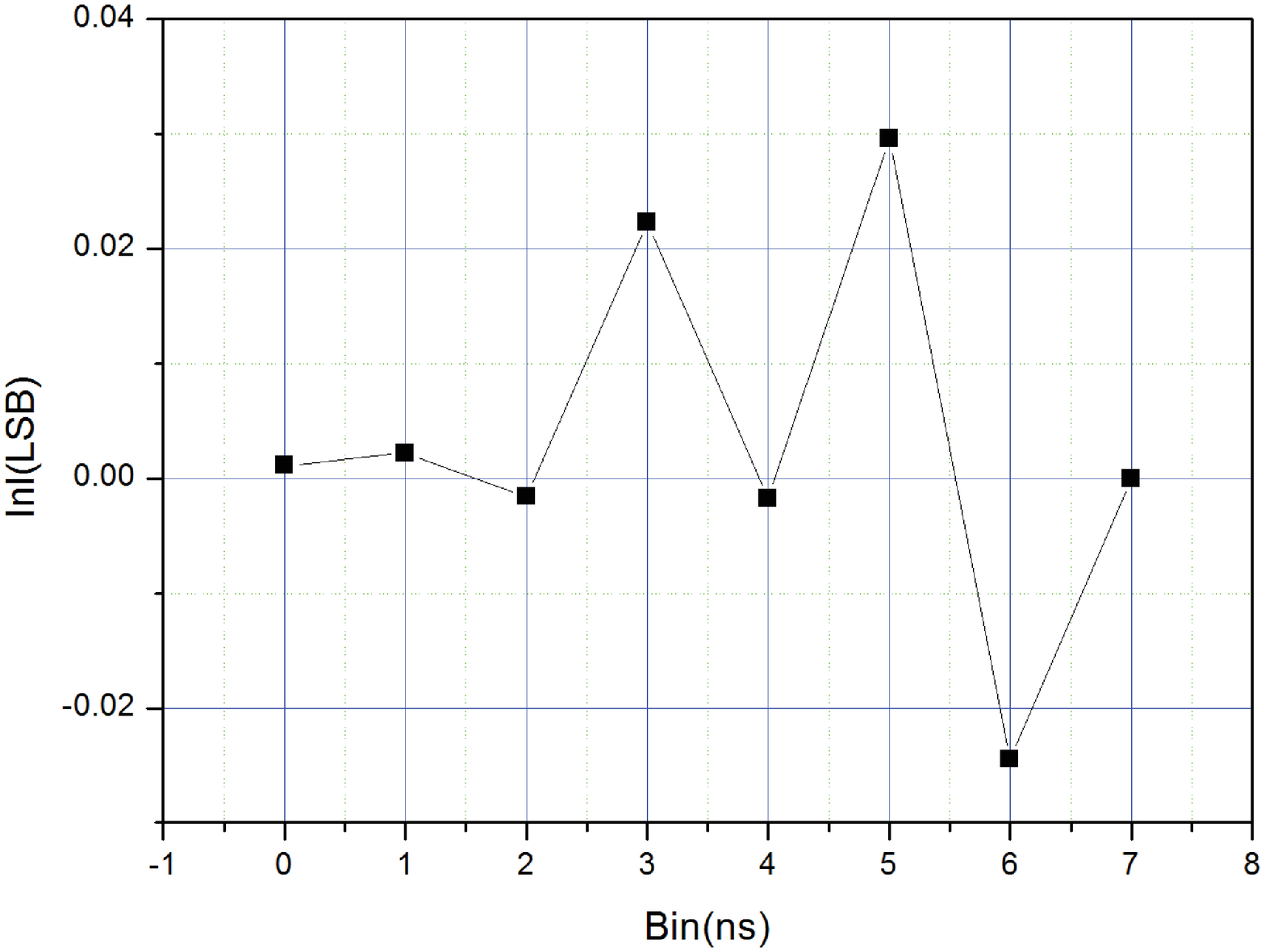}
\figcaption{\label{fig7}   INL of time digitizing module}
\end{center}

In the precision test, threshold is set at 1/8 particle. Fig.~\ref{fig8} shows the precision in different situations: the minimum input signals and the maximum input signals. Result indicates that the accuracy maintains better than 0.5ns no matter with the peak value of signals in laboratory.
\end{multicols}
\begin{center}
\includegraphics[width=12cm]{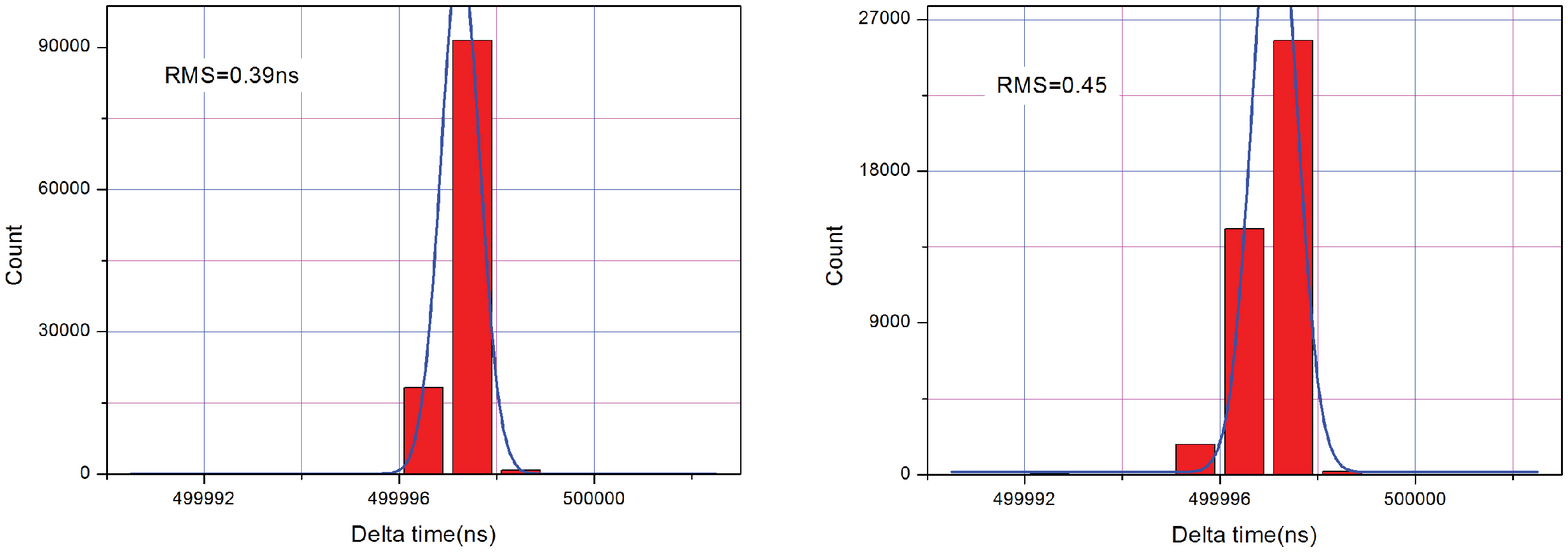}
\figcaption{\label{fig8}   Precision of time digitizing module in laboratory }
\end{center}
\begin{multicols}{2}

\begin{center}
\includegraphics[width=6cm]{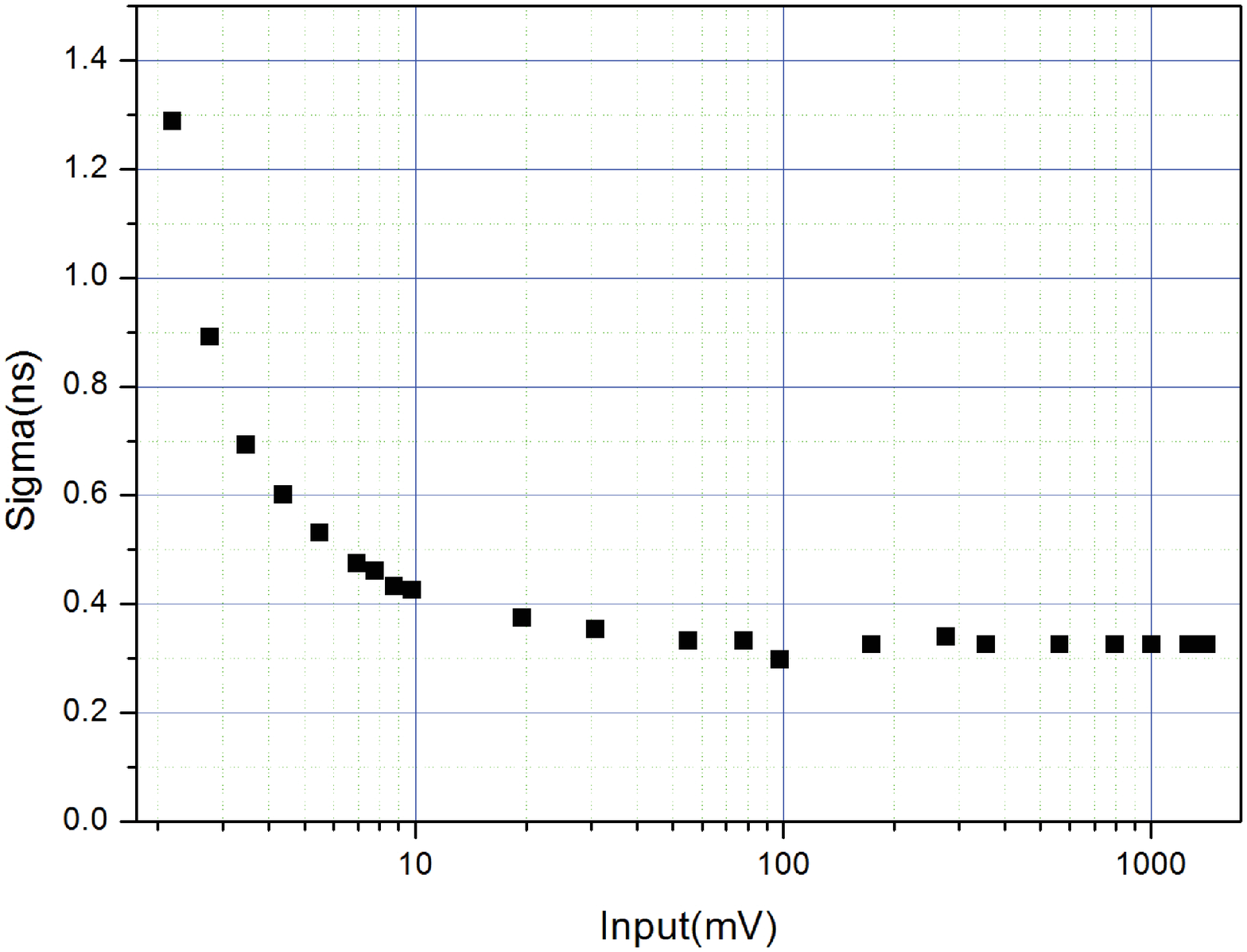}
\figcaption{\label{fig9}   Precision of time digitizing module with PMT }
\end{center}

However, experiment associating with PMTs shows that the precision gets worse when the input signal get smaller, as shown in Fig.~\ref{fig9}. The time-walking and noise of input signals contribute to the phenomenon that the accuracy is worse than 0.5ns when the peak value of signals get lower than 0.5 particle. In any case, this result has reached the target.

\subsection{Temperature performance}
As one of the key performances, the temperature effect is taken into special consideration in the system design. To assess this performance, we propose an experiment as shown in Fig.~\ref{fig9}. In this experiment, two synchronized readout boards are employed to test exactly same signals generated by the same generator. One of the two boards is placed under the normal condition to serve as a reference, while the other one is put in a thermostat, in which the temperature is programmed to varying from $-$10$^\circ$C to $+$60$^\circ$C.

\begin{center}
\includegraphics[width=6cm]{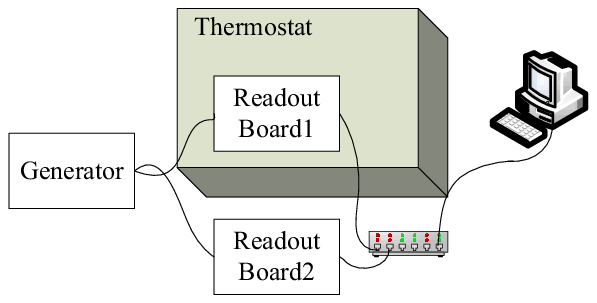}
\figcaption{\label{fig10}   Temperature performance experiment}
\end{center}
Then, the charge and time information of two boards in different situations reflect the temperature performance. The temperature of charge measurement is shown in fig.~\ref{fig11}. The Anode charge measurement presents a same trend with Dynode when temperature changes. The ADC output data has a tiny change in the whole temperature range while the input signals keep fixed. The ADC output varies less than 0.8\%\ in 70$^\circ$C, this characteristic has achieve the target because the PMT temperature coefficient is proved to be 0.2\%\ in 1$^\circ$C.
\begin{center}
\includegraphics[width=7cm]{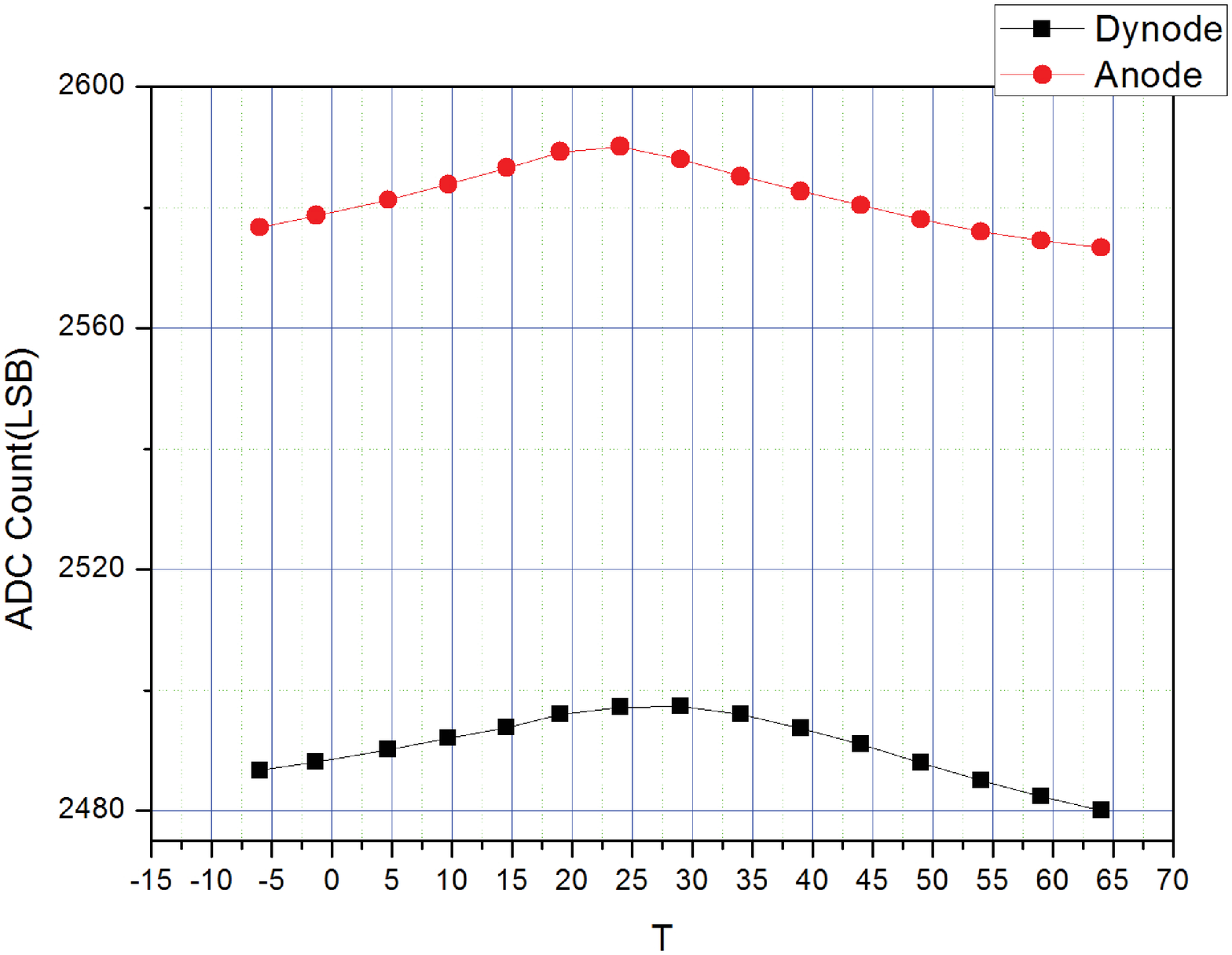}
\figcaption{\label{fig11}  Temperature performance of charge measurement}
\end{center}

As presented in fig.~\ref{fig12}, the arrival time tested by two boards keeps a 6ns delay because the different length of cables. This delay changes less than 150ps in 70$^\circ$C, which means that no correction is necessary in DAQ.
\begin{center}
\includegraphics[width=6cm]{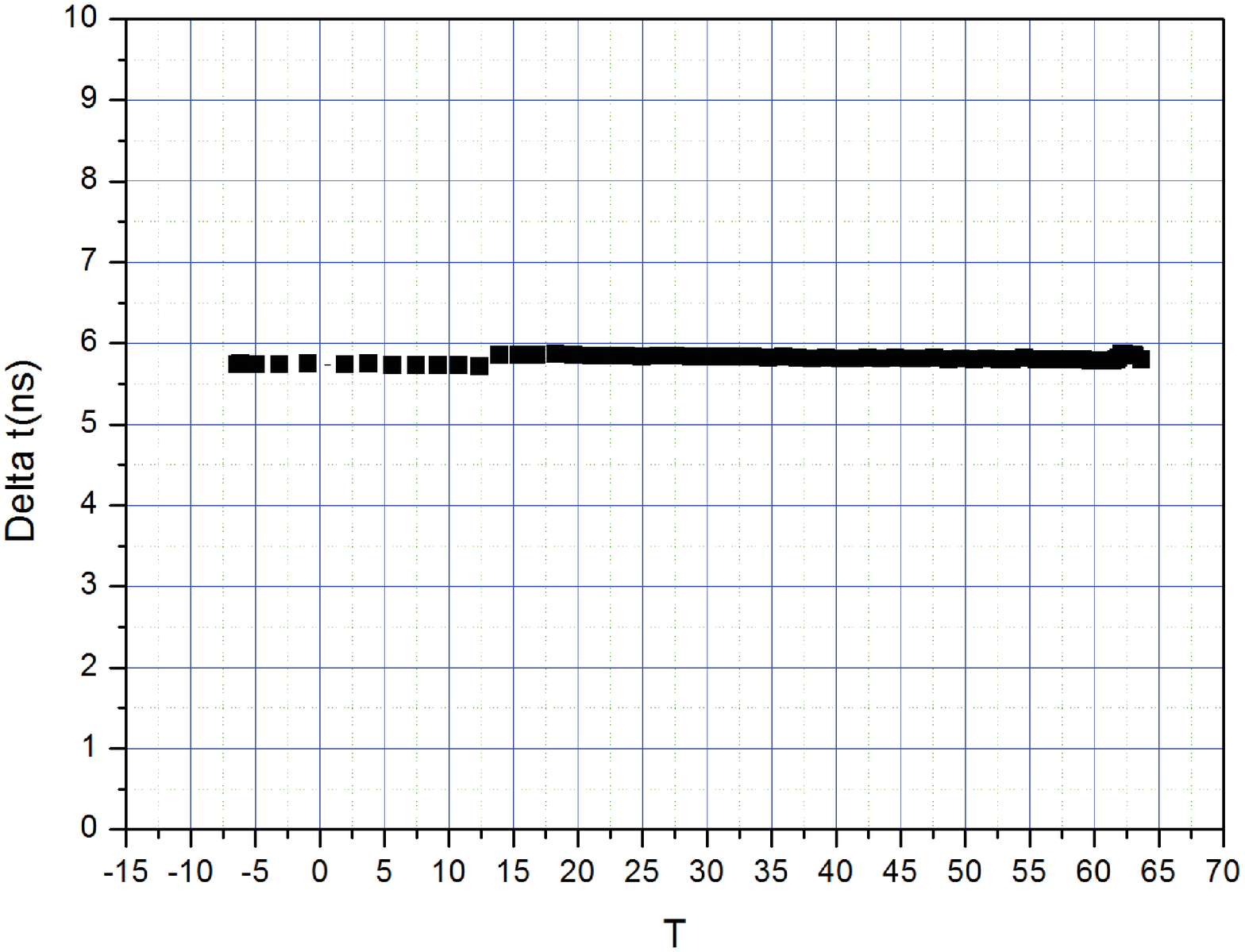}
\figcaption{\label{fig12}  Temperature performance of time measurement}
\end{center}

\section{Conclusion}
In this paper we propose a readout electronics system used for EDs in LHAASO KM2A. Each performances have meet the requirements and every index is proved to be stable in the large temperature range. With the technology adopted in this paper, readout electronics system can be designed for the high energy experiments built in the large, harsh environment.
\section{Footnotes}

\acknowledgments{We thank PAN Weibin and LI Hongming of Tsinghua University and LIU Jia of IHEP for their help.}

\end{multicols}

\vspace{10mm}

\begin{multicols}{2}

\end{multicols}

\vspace{-1mm}
\centerline{\rule{80mm}{0.1pt}}
\vspace{2mm}

\begin{multicols}{2}

\end{multicols}

\clearpage

\end{document}